\PassOptionsToPackage{dvipsnames}{xcolor}
\documentclass[conference]{IEEEtran}
\IEEEoverridecommandlockouts

\usepackage{cite}
\usepackage{amsmath,amssymb,amsfonts}
\usepackage{algorithmic}
\usepackage{graphicx}
\usepackage{textcomp}
\usepackage{float}
\usepackage{multirow}
\usepackage{array}
\usepackage{nicematrix}
\usepackage{caption}
\usepackage{subcaption}
\usepackage{soul}
\usepackage[hyphens]{url}

\usepackage{hyperref}
\usepackage{doi} 

\hypersetup{
  colorlinks=true,
  linkcolor=black,        
  citecolor=black,        
  urlcolor=blue,          
  breaklinks=true
}

\def\BibTeX{{\rm B\kern-.05em{\sc i\kern-.025em b}\kern-.08em
    T\kern-.1667em\lower.7ex\hbox{E}\kern-.125emX}}

\begin{document}

\title{Quantum Advantage in Computational Chemistry?


\thanks{E.C.~was supported in part by ARO MURI (award No.~SCON-00005095), and DoE (BNL contract No.~433702). FutureTech funding provided by Accenture.} }


\author{\IEEEauthorblockN{1\textsuperscript{st} Hans Gundlach}
\IEEEauthorblockA{\textit{FutureTech Lab} \\
\textit{MIT}\\
Cambridge, MA, USA \\
hansgund@mit.edu, \href{https://orcid.org/0000-0001-5499-5072}{ORCID}}
\and
\IEEEauthorblockN{2\textsuperscript{nd} Keeper	Sharkey}
\IEEEauthorblockA{\textit{ } 
\textit{ODE, L3C}\\
Sheridan, WY, USA \\
sharkey@odestar.com, \href{https://orcid.org/0000-0002-3767-6261}{ORCID}}
\and
\IEEEauthorblockN{3\textsuperscript{rd} Jayson Lynch}
\IEEEauthorblockA{\textit{FutureTech Lab} \\
\textit{MIT}\\
Cambridge, MA, USA \\
jaysonl@mit.edu}
\and
\IEEEauthorblockN{4\textsuperscript{th} Victoria Hazoglou}
\IEEEauthorblockA{\textit{Accenture Innovation} \\
\textit{Accenture}\\
New York City, NY, USA \\
victoria.perrone@accenture.com}
\and
\IEEEauthorblockN{5\textsuperscript{th} Kung-Chuan Hsu}
\IEEEauthorblockA{\textit{Accenture Innovation} \\
\textit{Accenture}\\
Los Angeles, CA, USA \\
kung-chuan.hsu@accenture.com}
\and
\IEEEauthorblockN{6\textsuperscript{th} Carl Dukatz}
\IEEEauthorblockA{\textit{Accenture Innovation} \\
\textit{Accenture}\\
Detroit, MI, USA \\
carl.m.dukatz@accenture.com}
\and
\IEEEauthorblockN{7\textsuperscript{th} Eleanor Crane}
\IEEEauthorblockA{\textit{MIT} \\
\textit{Belfer Center, Harvard Kennedy School}\\
Cambridge, USA \\
emc2@mit.edu, \href{https://orcid.org/0000-0002-2752-6462}{ORCID}}
\and
\IEEEauthorblockN{8\textsuperscript{th} Karin Walczyk}
\IEEEauthorblockA{\textit{Accenture Research} \\
\textit{Accenture}\\
D\"{u}sseldorf, Germany \\
karin.walczyk@accenture.com}
\and
\IEEEauthorblockN{9\textsuperscript{th} Marcin Bodziak}
\IEEEauthorblockA{\textit{Accenture Research} \\
\textit{Accenture}\\
Warsaw, Poland \\
marcin.bodziak@accenture.com}
\and
\IEEEauthorblockN{10\textsuperscript{th} Johannes Galatsanos-Dueck}
\IEEEauthorblockA{\textit{Initiative on the Digital Economy} \\
\textit{MIT}\\
Cambridge, MA, USA \\
galatsan@mit.edu}
\and
\IEEEauthorblockN{11\textsuperscript{th} Neil Thompson}
\IEEEauthorblockA{\textit{FutureTech Lab} \\
\textit{MIT}\\
Cambridge, MA, USA \\
neil\_t@mit.edu}
}
\maketitle

\begin{abstract}




%

For decades, computational chemistry has been posited as one of the areas in which quantum computing would revolutionize. However, the algorithmic advantages that fault-tolerant quantum computers have for chemistry can be overwhelmed by other disadvantages, such as error correction, processor speed, etc. To assess when quantum computing will be disruptive to computational chemistry, we compare a wide range of classical methods to quantum computational methods by extending the framework proposed by Choi, Moses, and Thompson~\cite{choi2023quantumtortoiseclassicalhare}. Our approach accounts for the characteristics of classical and quantum algorithms, and hardware, both today and as they improve.

We find that in many cases, classical computational chemistry methods will likely remain superior to quantum algorithms for at least the next couple of decades. Nevertheless, quantum computers are likely to make important contributions in two important areas. First, for simulations with tens or hundreds of atoms, highly accurate methods such as Full Configuration Interaction and high-order Coupled Cluster are likely to be surpassed by quantum phase estimation in the coming decade. Secondly, in cases where quantum phase estimation is most efficient, moderately accurate classical techniques, such as Møller–Plesset, could be surpassed in ten to fifteen years if the technical advancements for quantum computers are favorable. Overall, we find that in the next decade or so, quantum computing will be most impactful for highly accurate computations with small to medium-sized molecules, whereas classical computers will likely remain the typical choice for calculations of larger molecules. 
\end{abstract}

\section{Introduction}
Computational chemistry is of fundamental value to science, enabling enhanced understanding of chemical processes, and technological innovation within industry. Insights from computational chemistry predictions and verification of experimental measurements accelerate the in-silico discovery of novel drugs and materials.
Several of the most cited scientific papers of all time are related to computational chemistry \cite{natureTop100}. However, computational chemistry can be highly resource-intensive and time-consuming, often pushing the boundaries of existing computational hardware. It has been proposed that quantum computers could solve these computational bottlenecks and revolutionize the field. 
In particular, chemistry has been named the ``killer application of quantum computing" \cite{doi:10.1073/pnas.1619152114}.
Quantum computers were originally proposed as machines capable of effectively simulating quantum systems and have shown theoretical algorithmic advantage in domains such as cryptography and unstructured search \cite{choi2023quantumtortoiseclassicalhare}.  However, quantum computers come with engineering challenges, large financial overhead, and complexities that are not present in classical computing methods.

Given the trade-off between potentially enhanced algorithms on quantum computers and their possibly greater overhead compared to classical implementations, researchers have started to evaluate the practical value of quantum advantage. Drug design is a leading application of computational chemistry, and a growing body of literature is starting to evaluate the advantage of quantum computers in this area. However, many drug systems without strong correlations can already be treated with lower-accuracy methods \cite{santagati2024drug}. 
While this work focuses on ground-state energy estimation, other tasks such as real-time dynamics or excited states may see different quantum advantage timelines. Simulating dynamics classically faces challenges like entanglement growth, motivating further research on quantum algorithms for these tasks.

While many biological systems can be treated with lower-accuracy methods, numerous important drug complexes and enzymes exhibit strong electronic correlation, especially those with transition metal centers such as Cytochrome P450, which features an iron center \cite{Goings_2022}. In the context of biological systems and chemical use cases, an iron (Fe) and molybdemun (Mo) complex (FeMo cofactor), is a complex involved in nitrogen fixation, which has perceived market value if well understood computationally, leading to an increase in efficiency for ammonia production \cite{Reiher_2017}. This chemical system is a promising target for quantum advantage and a benchmark for quantum simulations \cite{günther2025phaseestimationpartiallyrandomized}, and it is traditionally very difficult to study with classical methods.  

Beyond drug design and nitrogen fixation, recent studies have assessed quantum computing across a broad range of chemical applications. These studies demonstrated advantages in certain high-accuracy scenarios, although the overall outcomes have been somewhat mixed. Bellonzi et al. \cite{bellonzi2024feasibilityacceleratinghomogeneouscatalyst} determined that quantum computers could be useful for homogeneous catalyst discovery with continued development. Agrawal et al. \cite{agrawal2024quantifyingfaulttolerantsimulation} found that advances would be necessary for simulating strongly correlated systems using the Fermi-Hubbard model. Nguyen et al.\cite{nguyen2024quantumcomputingcorrosionresistantmaterials} did a detailed study on quantum computers for corrosion-resistant materials showing utility if quantum computers are developed with thousands to hundreds of thousands of logical qubits. However, Otten et al. \cite{otten2024quantumresourcesrequiredbinding} found that the quantum resources necessary for modeling binding affinity of amyloid-beta to be too large for practical use.

In light of the importance of assessing the utility of quantum computers, this paper advances the state of the art in two key aspects. First, we take a broader perspective. Instead of limiting our evaluation to specific molecules or sets of molecules, we evaluate the utility of quantum computers against a diverse array of contemporary computational chemistry algorithms at various scales. As part of this study, we conduct an extensive survey of current classical and quantum methods in computational chemistry. Second, we go beyond evaluating quantum computing based solely on present-day hardware limitations, and instead we incorporate emerging trends in quantum hardware development. This allows us to evaluate the utility of quantum computers over time, aligning our analysis with anticipated advancements in quantum computers. 
Our model assumes significant classical parallelism (e.g., thousands of GPUs) while treating quantum algorithms as mostly serial. In the future, quantum architectures may also exploit parallelism—for example, via parallel Trotter steps or distributed fault-tolerant circuits—potentially shifting advantage timelines.

\section{Classical and Quantum Chemistry Algorithms}

\begin{table*}[t]
\centering
\large
\begin{NiceTabular}{ll|cc}[hvlines]

\textbf{Classical Method} & \textbf{Classical Time} & \rule{0pt}{2.6ex} \(\mathbf{O(N^3/\epsilon)}\) QPE &\(\mathbf{O(N^2/\epsilon)}  \)  QPE \\
\hline
Density Functional Theory  (DFT)  & \(O(N^{3})\)     &  \newline     
                                    N/A & \newline \(>2050\) \\
Hartree Fock (HF)       & \(O(N^{4})\)     &  \(>2050\) \newline 
                                    & \newline 2044 \\
M\o ller-Plesset Second Order (MP2)      & \(O(N^{5})\)     & \(>2050\) \newline 
                                    &  \newline 2038 \\
Couple Cluster Singles and Doubles (CCSD)     & \(O(N^{6})\)     &  2044 \newline
                                    &  \newline 2036 \\
\shortstack{Couple Cluster Singles and Doubles \\ Perturbative Triples (CCSD(T))} & \(O(N^{7})\)     &  2036 \newline 
                                    &  \newline 2034 \\
Full Configuration Interaction (FCI)     & \(O^*(4^{N})\)   & 2032 \newline 
                                    &  \newline 2031 \\
\end{NiceTabular}

\caption{Comparison of classical and quantum (QPE) methods with their asymptotic times and estimated first year of quantum economic advantage. The N/A gives cases where QEA is never achieved because there is no algorithmic advantage. The error tolerance is set to $\epsilon=10^{-3}$, $N$ represents the number of relevant basis functions \cite{low2025fastquantumsimulationelectronic}. Discussion on how we handle these comparisons can be found in Appendix~\ref{sec:Orbital to Atom Ratio}.}

\label{tab:chemistry_methods}
\end{table*}

\subsection{Classical Algorithms for Computational Chemistry}

Quantum computational chemistry is a large and diverse field. In order to evaluate the utility of quantum algorithms, we must first understand the breadth of classical algorithms available. Our work primarily analyzes post-Hartree Fock methods like Coupled Cluster and Full Configuration Interaction as these have the most potential for quantum disruption. However, algorithms like Density Functional Theory are useful points of comparison in understanding the current quantum chemistry landscape. For many quantum chemistry algorithms like Coupled Cluster Singles and Doubles or Hartree Fock, there exist many different asymptotic times depending on specific approximations used. For each algorithm, we select a representative time complexity that accurately captures its performance characteristics. We have chosen comparisons across a wide range of time complexities. This approach enables the readers to compare disruption points across various complexity classes, although our chosen complexities do not necessarily correspond to the most optimized implementations possible.

\subsubsection{Density Functional Theory}


Density functional theory (DFT) is one of the most popular methods. This method uses electronic density in place of electronic wave functions to determine the ground state energy. DFT has remarkable asymptotic scaling with some implementations for large system scaling $O(N)$ in the number of basis functions \cite{nguyen2024quantumcomputingcorrosionresistantmaterials}. However, DFT generally scales cubically with the number of atoms $O(N_{at}^3)$ \cite{ko2023implementation}. 
While some Gaussian basis sets lead to approximate proportionality of basis functions to atoms, in practice the choice of basis set significantly affects scaling. For example, pure functionals like LDA scale as  $O(N^3)$, but hybrid functionals including Hartree-Fock exchange dominate scaling at $O(N^4)$ \cite{mohr2015accurate}.
Furthermore, in practice, DFT depends on an approximation of the exchange-correlation functional and there is no one-size-fits-all way to perform this approximation \cite{nguyen2024quantumcomputingcorrosionresistantmaterials}.
Moreover, for molecules, plane-wave bases are rarely used and localized basis sets introduce significant variations in scaling.




\subsubsection{Hartree Fock}

Hartree Fock (HF) was one of the first methods developed in quantum chemistry. HF is considered a low-accuracy method due to its lack of correlation energies \cite{Field2017}. However, HF is used as a starting point approximation for more accurate methods such as the M\o ller-Plesset (MP) perturbation theory. Conventionally, Hartree Fock scales with the number of basis functions as $O(N^{4})$ \cite{Loos2022_HFPostHF}. This is primarily due to the four-index integrals that describe electron repulsion. However, new methods can significantly reduce this asymptotic time. This includes density-fitting approaches that can lower the scaling to $O(N^{3})$ \cite{Field2017}. Other approaches include employing Cholesky decomposition techniques, as well as methods utilizing localized molecular orbitals, which can significantly reduce the scaling of Hartree–Fock calculations for larger systems \cite{goletto2021linear}.

\subsubsection{Post Hartree Fock Methods}
MP methods include perturbations of arbitrary order (e.g., MP2, MP3, and higher), though in practice low-order methods like MP2 are most commonly used.
In this work, MP2 is the focus of the analysis. MP2 scales with $O(N^{5})$ due to the two-electron integrals with a four-index summation. 

In chemical systems where higher accuracy is needed to represent a relevant measurement,  Coupled Cluster (CC) methods are employed.  CC methods use an efficient wave function ansatz based on the exponentiation of the cluster operator. CC Singles and Doubles (CCSD), meaning single and double excitations, is the most common form of this method. In chemical applications requiring even greater accuracy, CC methods incorporating triple excitations, such as CCSD with full triples (CCSDT) or perturbative triples (CCSD(T)), are employed. 
CCSD(T) has been termed the ‘gold standard’ for many molecular systems, sometimes achieving accuracies comparable to experimental measurements of molecular properties.

These methods have a higher-order complexity as compared to the prior methods discussed. 
CCSD scales as $O (N^{6})$), reflecting the cost of treating single and double excitations. CCSD(T) achieves approximate inclusion of triple excitations with $O (N^{7})$ scaling by selectively including dominant contributions (rather than all triple excitation terms, as in CCSDT's $O (N^{8})$ scaling).
The growing complexity associated with the excitations is the limiting factor for using higher-order CC methods in practice. CCSD(T) even further tries to alleviate the complexity by an order of magnitude to $O(N^{7})$\cite{hohenstein2022rank}, by not including the full range of triple excitations. As with all of these methods there are numerious variations and adaptations in implimentation details, for example \cite{watts1993coupled, raghavachari1989fifth, urban1985towards}.

Full Configuration Interaction (FCI) is the most accurate method in computational chemistry, although it still relies on certain approximations of the wave function—most notably, the Born–Oppenheimer approximation, which assumes a significant mass difference between electrons and nuclei.
Nonetheless, FCI scales as the binomial coefficient dependent on the number of spin orbitals ($N_{so} =2N$) as well as $N_{e}$; $O(b(N_{so}, N_{e}))$ \cite{Field2017}, which is impractical for most all chemically relevant use cases. 

Complete Active Space (CAS) alleviates the scaling issue by only running FCI on a subset of the most important orbitals known as the ``the active space." CAS scales like FCI in terms of the active space, and since it is hard to determine the active space necessary to achieve a chemically relevant prediction for many use cases we neglect explicit analysis of CAS algorithms in this analysis. 

In our analysis, we need to find an asymptotic form for FCI, which depends only on the number of basis functions ($N$) rather than electrons $N_{e}$.
We assume a half-filled system for estimation, where each spatial orbital is on average singly occupied, though in practice spin-paired electrons can share the same spatial orbital.
In other words, the orbitals only have one electron, however, two spin-paired electrons are physically allowed in each orbital. The binomial coefficient $\tbinom{N_{so}}{N_{e}}=\tbinom{N_{so}}{N_{so}/2}$ is maximal for fixed $N_{so}$ which gives a good upper bound for the runtime of FCI. Moreover, many strongly correlated systems such as transition metal complexes are effectively half-filled in the outer shells or valence orbitals; core electrons are in spin-paired orbitals. Using the Sterling Approximation the complexity for FCI leads to $O^*(2^{N_{so}})$ scaling or $O^*(4^{N})$ where $N$ is the number of spatial orbital basis functions, where $O^*$ notation neglects polynomial factors in $N$.

\subsubsection{DMRG, VMC, and other classical methods}
There is an enormous range of classical methods in quantum chemistry. Our analysis is primarily focused on general purpose chemistry algorithms with high asymptotic scaling i.e at least $\Omega(N^4)$. However, we would be remiss not to mention other popular techniques. 
DMRG (Density Matrix Renomalization Group) is a method growing in popularity and is the method of choice for one-dimensional quantum systems \cite{schollwock2005density}. DMRG is based on the construction of a wavefunction ansatz called the Matrix Product State (MPS) \cite{nguyen2024quantumcomputingcorrosionresistantmaterials}. DMRG has an asymptotic time of $O(k^3 M^3)$ where $k$ is the number of active orbitals and $M$ is the bond dimensions which controls the answer quality \cite{Goings_2022}.
Another promising direction in classical computational chemistry methods are Monte-Carlo based algorithms.
Quantum variational Monte-Carlo (VMC) methods try to minimize an energy functional $E[\phi_{N}(\theta)]$ where $\phi_{N}(\theta)$ is an N-particle wavefunction approximation parametrized by $\theta$. One can then use Metropolis-Hasting algorithm to sample a configuration from $\phi_{N}(\theta)$ and use Monte-Carlo integration to determine its energy. VMC usually scales between $O(N_{e}^3)$ and $O(N_{e}^4)$ \cite{nguyen2024quantumcomputingcorrosionresistantmaterials}.
Our analytical framework is not suited to address AI-based methods for computational chemistry in either classical or quantum computers. We include a short discussion about this in Sec.~\ref{Limitations and Further Work}.

\subsection{Quantum Chemistry Algorithms on Quantum Computers}
\label{sec:Quantum Chemistry Algorithms on Quantum Computers}

There exist many approaches to use quantum computers for solving problems in quantum chemistry. We focus on fault-tolerant quantum algorithms, in particular, quantum phase estimation (QPE). We exclude variational approaches due to unclear asymptotic times. However, QPE encompasses many distinct implementations and variations, and there has been rapid progress in T-gate counts for the algorithm. Early QPE implementations had T-gate scaling of $O(N^{10}/\epsilon^{3/2})$ \cite{Lee_2021}. Later advancement reduced T-gate scaling to $O(N^{5}/\epsilon)$ \cite{Lee_2021}. Furthermore, QPE asymptotic times depend on the basis type. Using a plane-wave basis leads to highly structured sparse Hamiltonian operators which scale ~$\tilde{O}(N^{3}/\epsilon)$ or even $\tilde{O}{(N^{2}/\epsilon)}$ (this latter method has high logical qubit overhead)\cite{Babbush_2019}. The plane-wave basis is most often used for materials or uniform electron gas models. However, for localized molecules, a very large number of basis functions $N$ is necessary for chemical accuracy \cite{Berry_2019}. In this case, first-quantization based QPE approaches can be used. If first-quantization methods are used, logical qubits scale as $O(N_{e}  log(N))$ and T-gates scale as $O(N_{e}^{8/3}N^{1/3})$, where $N_{e}$ is the number of electrons and $N$ is the number of basis functions \cite{Babbush_2019}.

As a caveat, asymptotic times for Hamiltonian simulation and QPE are somewhat misleading. The problem of finding the ground state is QMA-complete in cases like the k-local Hamiltonian problem \cite{kempe2005complexitylocalhamiltonianproblem} when $k \leq 3$. QPE depends on developing an initial state that is sufficiently close to the final state. In practice, most research in QPE neglects this or assumes that techniques like Hartree-Fock are sufficient to prepare the initial state \cite{fomichev2024initial}. The T-gate overhead due to real-world state preparation and imperfect initial state for FeMoco has been estimated to add a factor of 2 compared to estimates without this consideration \cite{berry2024rapid}. 
Initial state preparation remains a major practical challenge, particularly for strongly correlated systems like FeMoco. While we assume efficient preparation for this study, real-world performance may incur additional overhead without reliable methods for preparing high-overlap initial state. We note that our assumptions of $O(N^3/\epsilon)$ scaling may be optimistic for some Hamiltonian simulation methods. Varying this scaling affects predicted timelines significantly, and future work could further refine these assumptions.

Moreover, let $F$ be the fidelity between the initial state and the true eigenstate; the average number of times the QPE circuit needs to be repeated scales as $O(1/F)$ \cite{Ge_2019}. Therefore, careful preparation of the initial state is essential to avoid the substantial cost of re-running the QPE circuit. In the recent decade, much effort has been made in the research on efficient state preparation to increase $F$. Configuration state functions (CSF) methods~\cite{Babbush_2015,Sugisaki_2016,Sugisaki_2019,Tubman_2018} provide a richer spectrum of wavefunction, which could can significantly enhance the accuracy of initial state overlap. Matrix-product states (MPS) methods~\cite{Malz_2024} also provides the needed flexibility that can be used to efficiently model initial states in strongly correlated quantum systems. Adiabatic state preparation~\cite{Aspuru-Guzik_2005} is also a natural candidate to evolve a ground state.

Quantum phase estimation is based on estimating the phase resulting from Hamiltonian simulation. Hamiltonian simulation focuses on approximately applying a unitary time-evolution operator $U = e^{-iHt}$ of a hamiltonian $H$ for some time $t$. The simplest QPE approaches using Trotter-Suzuki Product formulas generally have $\widetilde{O}(N^5/\epsilon)$ scaling \cite{babbush2016exponentially}. Recent qubitization based approaches generally have significantly improved asymptotic times with $\widetilde{O}(N^{3/2}\lambda/\epsilon)$ T-gate complexity and has $\widetilde O(N^3)$ scaling when $\lambda= \Omega(N^{3/2})$, which is generally the case \cite{Berry_2019}. Tensor-hypercontraction (THC) along with qubitization has lead to  $\tilde{O}(N \lambda_{\zeta}/\epsilon)$ scaling with $\lambda_{\zeta}$ scaling between $O(N)$ and $O(N^3)$ \cite{Lee_2021}. In particular, Lee et al. \cite{Lee_2021} demonstrated an asymptotic scaling of $\widetilde{O}(N^{2.1}/\epsilon)$ for the QPE-based method in the thermodynamic limit of the hydrogen chain. In this paper, we choose cubic scaling for QPE as our main estimate for the asymptotic scaling of QPE. We chose this estimate based on the range of implementations that have this scaling. Quadratic QPE represents an optimistic view, which is true for some systems and may become more generally the case with further algorithmic progress. 
$\widetilde O(N^5)$ represents more of the historical case for comparison.

\subsubsection{Beyond QPE Methods}
QPE is not the sole algorithm used for quantum chemistry on quantum computers. However, it is generally more widely studied. More importantly for our case, it has relatively clear asymptotic bounds. 
Quantum variational algorithms are another popular technique that might hold great promise \cite{wu2024variational}. Nevertheless, it is very difficult to characterize their asymptotic behavior. For instance, McClean et al. showed that variational quantum circuits become significantly harder to train at larger scales due to the phenomena of "Barren Plateaus" \cite{mcclean2018barren}.

Other approaches for determining ground state energies in quantum chemistry include adiabatic passage techniques \cite{Granet_2025}, imaginary time evolution \cite{motta2020determining}, subspace, and Lanczos methods \cite{parrish2019quantum,kirby2023exact}. These approaches may also be promising but are less well studied and characterized than either QPE or variational approaches. 

\subsection{Quantum Advantage Framework}\label{quantum_advantage_framework}

\begin{figure*}[!t]
\centering
\begin{minipage}{0.45\textwidth}
  \centering
  \includegraphics[width=\linewidth]{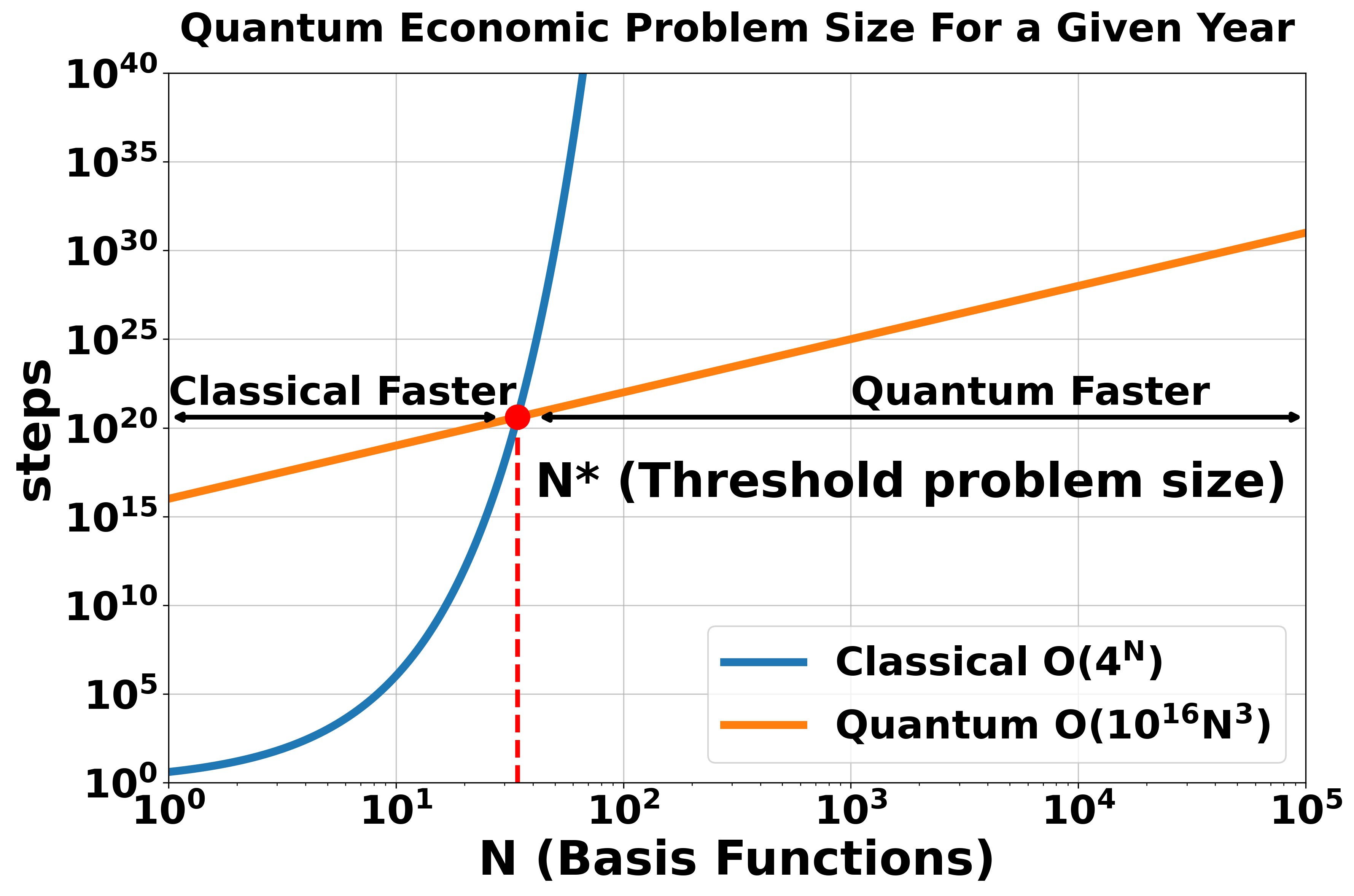}
  \caption{Quantum Economic Advantage (QEA) Problem Size in a given year. Quantum computers come with large constant hardware overheads but algorithmic advantages. Therefore, problem sizes must be over a critical threshold to have an advantage on a quantum computer.}
  \label{fig: qea illustration}
\end{minipage}%
\hfill
\begin{minipage}{0.5\textwidth}
  \centering
  \includegraphics[width=\linewidth]{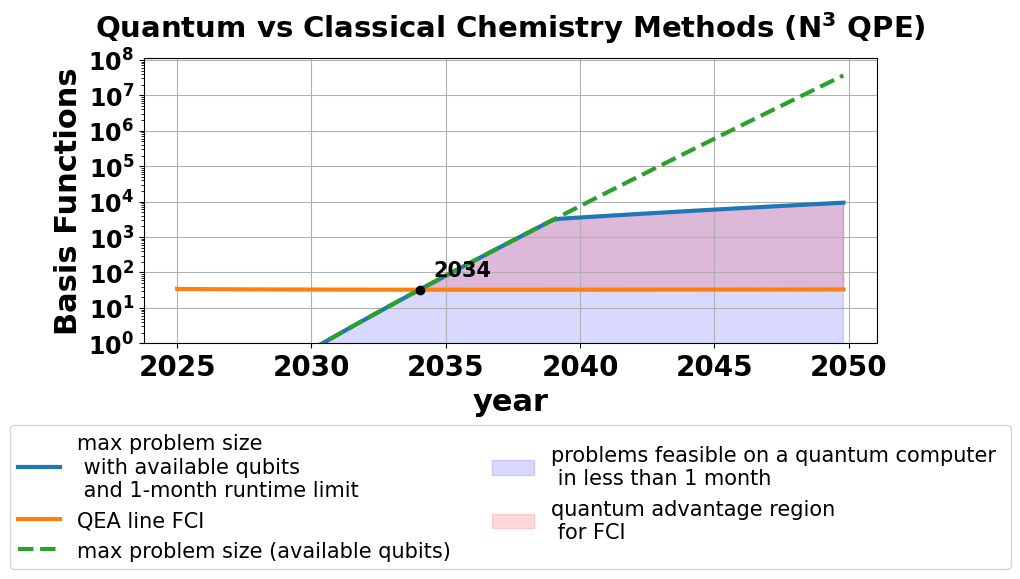}
  \caption{QEA threshold problem size over time taking into account hardware trends in quantum-classical overhead. The green line represents the maximum theoretical size that could be done without any time limit. We also choose to include a time limit of 1-month to better model constraints for computational chemistry. The quantum advantage region illustrates problem sizes that are feasible and preferable to run on a quantum computer in relation to a given classical algorithm.}
  \label{fig:generalized qea illustration}
\end{minipage}
\end{figure*}

In order to evaluate the practical utility of quantum algorithms over classical algorithms for quantum chemistry we extend the framework developed by Choi, Moses, and Thompson \cite{choi2023quantumtortoiseclassicalhare}. Recent work has also extended this work to include classical parallelism and more detailed hardware analysis \cite{MITQECalc, QEA}. This framework relies on the realization that problem sizes need to be sufficiently large to yield advantage for a a quantum computer. This is due to the fact that quantum computers come with significant overhead. We estimate on the order of a $10^{13}$ slowdown on price comparable hardware  (excluding algorithmic constants and $(1/\epsilon)$ overhead see Appendix: \ref{How Fast Are Quantum Computers?} for cost calculations and Tables~\ref{tab:classical_benchmark} and \ref{tab:Logical qubit and T-gate estimats for QPE} for estimates of algorithmic constants which appear comparable between classical and quantum algorithms). Even with strong algorithmic advantages, it is hard for quantum computers to overcome this barrier. For example, take the comparison between $O(N^3)$ QPE and $O(N^6)$ CCSD. For QPE to be more efficient than CCSD we must have a problem size such that roughly $10^{13}N^3 < N^6$ which is true for $N>10^{13/3}$.  We refer to the threshold problem size as the quantum economic advantage problem size.  
However, this quantum overhead relative to classical processors is rapidly changing. 

Advances in quantum gate fidelity, quantum error correction, and superconducting gate times could decrease this overhead. On the other hand, this overhead could increase if GPU progress or other classical processor progress increases at a faster rate. Due to these considerations, the quantum economic advantage size changes with time. The necessary size for quantum advantage is illustrated by the quantum economics advantage or QEA line in Fig ~\ref{fig:generalized qea illustration}. We choose to compare to superconducting qubit technology because superconducting qubits are one of the most popular quantum computing hardware paradigms and there is significant work evaluating the overhead costs and resource estimate of superconducting quantum computer \cite{Babbush_2021}\cite{sevilla2020forecastingtimelinesquantumcomputing}. Second, superconducting qubits have much faster gate and computation speeds than other quantum computing paradigms like trapped ion quantum computing (even taking into account error correction) \cite{Gschwendtner2023QuantumHardware}. Gate speed is an important determiner of quantum advantage. Large gate overheads makes it much harder to run practical problems on quantum computer in a reasonable time-frame (e.g. $<1$ month). However, superconducting quantum computers have lower qubit fidelity and generally require greater error correction overhead \cite{Gschwendtner2023QuantumHardware}. Quantum error correction adds significant overhead to quantum calculations. For instance, to represent one logical qubit takes on the order of hundreds to thousands of physical qubits \cite{choi2023quantumtortoiseclassicalhare}. In addition, implementing logical T-gates requires significant overhead due to the necessity for magic state distillation \cite{Babbush_2021}. Since T-gate implementation is much more resource intensive the number of T-gates required for algorithms is used as the most relevant cost estimate. In addition, quantum computers generally come with much greater costs per gate operation compared to classical floating point operations. This means price-comparable hardware can be much faster than naive operation overhead might suggest (see Appendix ~\ref{How Fast Are Quantum Computers?}). Taking all these considerations into account, we model the error correction overhead using surface code error correction and use trends in gate fidelity to infer the future physical to logical qubit ratio. We then use this ratio to infer gate-overheads and price overhead improvements. For classical computers, we assume increases in performance at the rate of Moore's law of about $40 \%$ per year.  Our estimates factor in growth from parallelism since the clock speed for individual processors has relatively stalled in recent years\cite{rupp2018_42years}. The detailed estimates for these improvements in quantum hardware and our model of quantum overhead we include in forthcoming work \cite{gundlach2024quantum_working}.

Even if a quantum algorithm and problem size have "economic" advantage relative to a classical algorithm. It still may not be feasible to run a quantum computer for this problem size, given two key limitations. First, there might not be enough qubits available to run the algorithm at that size. The variations of QPE in this paper require $O(N)$ qubits where N is the number of basis functions. We model qubit availability by exponentially extrapolating the roadmaps of quantum computer providers. We focus on IBM's roadmap\cite{IBM2024QuantumRoadmap} as this is a somewhat conservative roadmap in our study. It has a growth rate representative of the growth of top-end superconducting processors and is of relatively consistent quality. Extrapolating based on the maximum number of qubits often overestimates growth by including systems that sacrifice the number of qubits for qubit quality. For instance, in current systems larger number of qubits are generally correlated with less qubit fidelity and, therefore, much larger error correction overhead \cite{sevilla2020forecastingtimelinesquantumcomputing}.

In addition, we determine feasibility by limiting the processing time necessary to do the calculation on a quantum computer to one month. Computational chemistry methods can be time-consuming. For most use cases, such as chemical search and experimental interpretation, it is still necessary for simulations to finish in less than a month or a shorter time scale to be practically useful.

\subsubsection{Quantum Chemistry Algorithmic Constants}\label{Constants_Discussion}

A key constraint of our model is uncertainty in algorithmic constants. We have tried to address this in two principled ways. First, we conduct a robustness study where we vary the algorithmic overhead on quantum speed, classical speed, as well as qubit requirements. In large parts, our conclusions are robust to these changes. Especially to variation in classical constants \ref{Variation and Robustness Studies}. Second, we have surveyed the literature on algorithmic constants on both the quantum and classical chemistry algorithms. In general, classical time estimates look like between $.1-1$ times what we would expect using naive estimates based on the asymptotic formula with algorithmic constants equal to 1 and assuming GPU max performance for CCSD and CCSD(T) \ref{tab:classical_benchmark}. 
We don't have sufficient information for other methods like HF, MP2, and FCI. Given the small algorithmic constants in CC methods, we assume classical operations can be adequately estimated using basic asymptotic expressions.
However, as stated before our time estimates are robust to classical constants. Currently, there are no large-scale fault-tolerant quantum computers which makes it difficult to estimate algorithmic constants for quantum computers. Further, algorithmic constants on quantum methods have reduced significantly over time. Over the past few years, resource requirements (T-gates) for QPE have dropped by seven orders of magnitude \cite{low2025fastquantumsimulationelectronic}. We suspect that constant terms will decrease even more dramatically with the development of large-scale fault-tolerant quantum computers. Nevertheless, there have been several papers that try to make these resource estimates for chemistry problems \cite{otten2024quantum} \cite{nguyen2024quantumcomputingcorrosionresistantmaterials}. In general, we find the number of T gates in these studies can be well estimated using asymptotic expressions to within an order of magnitude (see Sec~\ref{tab:Logical qubit and T-gate estimats for QPE}). We assume in our estimate an accuracy level of $\epsilon \approx 10^{-3}$, which is the convention in other papers \cite{babbush2018encoding}\cite{günther2025phaseestimationpartiallyrandomized}. We also assume, as is the convention in many papers, that initial state preparation can be done relatively efficiently, and the initial state strongly overlaps with the ground state \cite{günther2025phaseestimationpartiallyrandomized}. Our results are mostly robust to increased overhead of about an order of magnitude (see Sec~\ref{Variation and Robustness Studies}). Otten et al. \cite{otten2024quantum} and Nguyen et al. \cite{nguyen2024quantumcomputingcorrosionresistantmaterials} also include estimates for required logical qubits. Overall, we see that the number of logical qubits required in these estimates is about 10x the number of basis functions. We use this tenfold overhead as our default in our logical qubit estimates. However, we include estimates without this constant logical overhead in our robustness analysis (Sec~\ref{Variation and Robustness Studies}). It is often possible to trade off increased qubit requirements for decreased Toffoli count and vice versa \cite{low2025fastquantumsimulationelectronic}\cite{günther2025phaseestimationpartiallyrandomized}. We do not believe that these classical and quantum benchmarks necessary capture all variations. However, they give some empirical grounding to our estimates.

\section{Results of Our Modeling}
\begin{figure*}[!t]
    \centering
    \includegraphics[width=\textwidth]{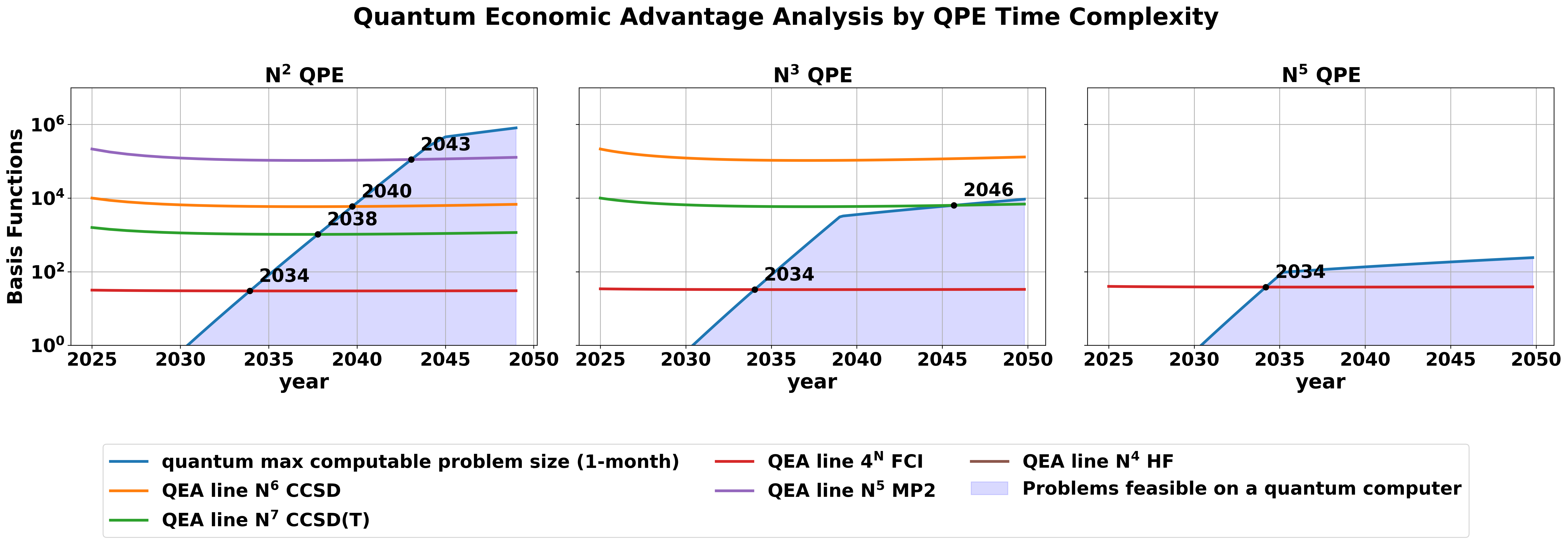}
    \caption{Evaluation of when quantum computing could replace select quantum chemistry algorithms at certain sizes. The region above each QEA line but below the maximum feasible problem size are problem sizes that are both feasible and less resource intensive to do on a quantum computer.}
    \label{fig:qea_chemistry_graph}
\end{figure*}

In this section, we examine when quantum computers are likely to disrupt classical simulation methods in quantum chemistry based on the quantum advantage framework. Figure~\ref{fig:qea_chemistry_graph} shows the plot for the problems that are feasible and advantageous. The first year of quantum advantage for each method is given in Table~\ref{tab:chemistry_methods}. Details on our model assumptions can be found in Section~\ref{quantum_advantage_framework}.  We see that our prediction for which algorithms will be replaced for some problems is determined by the asymptotic time of QPE. Yet, the year for disruption for some algorithms like FCI or CCSD(T) varies little with time complexity. In general, we embrace predictions with $O(N^3)$ QPE as our most confident predictions \ref{sec:Quantum Chemistry Algorithms on Quantum Computers}. 
For quantum algorithms that run in $O(N^3)$ time and $O(N)$ qubits where $N$ is the number of basis functions, we see chemistry simulations becoming possible on quantum computers in the early 2030s. Yet, we do not see quantum economic advantage until the mid-2030's. Further in the near future, we only see quantum computing replacing the most costly methods of quantum chemistry simulation, namely Full Configuration Interaction and Coupled Cluster with perturbative triplets. For systems with $N^2$ QPE scaling, all Post-Hartree Fock methods will see disruption in the 2030s and possibly some forms of Hartree Fock. In addition, for $N^2$ QPE in the 2040s, we might see quite large systems on the order of $10^5$ atoms being modeled on a quantum computer in less than a month. On the other hand, if we use older QPE implementations with $O(N^5)$ scaling, we only see FCI disrupted with limited systems sizes, mostly less than 10 atoms. FCI is not widely used in commercial quantum chemistry. For instance, by 2024, the largest FCI calculation ever done determined the exact energy of propane containing 26 electrons and 23 orbitals using the minimal STO-3G basis set \cite{loos2024greenselectedconfigurationinteraction}. 
Given the limited practical use of large-scale FCI calculations today, quantum computing’s earliest advantages may be confined to niche, high-accuracy scenarios rather than broadly transforming all of computational chemistry.
The difference in these scenarios emphasizes the important role algorithmic progress plays in the viability of quantum computing.

\subsection{Variation and Robustness Studies}\label{Variation and Robustness Studies}
We have tried to cross-check the constants used in our model. However, there are  a large variety of QPE implementations, with some implementations requiring more Toffoli gates in exchange for fewer qubits and vice versa. There may also be wide variations in classical constants depending on algorithmic implementation. 
Classical variation could include different levels of GPU utilization. Quantum computers might also have different levels of state initialization overhead. It is also interesting to incorporate further progress in quantum algorithmic constants. 
We conduct a range of variations in our model. In the first case, we examine the effect of decreasing the required logical qubit numbers by a factor of $10^{-1}$ , which might be a viable approach at reasonable Toffoli levels \cite{günther2025phaseestimationpartiallyrandomized}. This has a moderate 1-2 year effect on the date of disruption for many of the advances we study. The small effect of such a large qubit change is due to rapid growth in the number of physical qubits and decreases in the error correction overhead necessary to run these systems over time.   

In the second variation, we 10x the algorithmic constant for quantum algorithms, which is representative of Toffoli requirements of older approaches \cite{low2025fastquantumsimulationelectronic}. Adjusting the time constants has a moderate effect on the date of disruption for many algorithms. However, for some cases, the date of disruption is highly sensitive. This is in part due to our slow estimates of quantum speed improvements as well as the fast pace of classical computing progress. 
If a quantum algorithm cannot achieve advantage within a practical time frame (e.g., one month), even significant improvements in speed (e.g., a 10x increase) may not suffice to make it economically viable for routine use in the near term.

Finally, we make classical algorithms $10^3$ times faster (i.e., add a constant overhead of $10^{-3}$). This variation shows the robustness of our model to changes in classical constants as well as incorporate the effects of technologies witch might speed up classical chemistry by large constant factors. Such a large change has only a moderate effect because additional classical speed is relatively insignificant to the baseline overhead $10^{13}$, and such an improvement does not affect the one-month quantum limit. 


\begin{table}[!t]
    \centering
    \caption{\(N^3\) QPE robustness}
    \label{tab:n3-qpe-robustness}
    \resizebox{\columnwidth}{!}{
    \begin{tabular}{llccc}
        \hline
         & original & \(\times 10^{-1}\) logical & \(\times 10\) quantum-time & \(\times 10^{-3}\) classical-time \\
        \hline
        $N^4$ HF      &   $>2050$       & $>2050$ & $>2050$ & $>2050$ \\
        $N^5$ MP2     &$>2050$          & $>2050$ & $>2050$ & $>2050$ \\
        $N^6$ CCSD    &2044          & 2044    & $>2050$ & $>2050$ \\
        $N^7$ CCSD(T) &2036          & 2034    & 2036    & 2039 \\
        $4^N$ FCI     &2032          & 2029    & 2032    & 2032 \\
        \hline
        \vspace{1pt}
    \end{tabular}
    }
    \footnotesize{Variation of our predicted year of disruption based on changes in logical qubit requirements and quantum or classical constants.}
\end{table}

\begin{table}[!t]
    \centering
    \caption{\(N^2\) QPE robustness}
    \label{tab:n2-qpe-robustness}
    \resizebox{\columnwidth}{!}{
    \begin{tabular}{llccc}
        \hline
         & original & \(\times 10^{-1}\) logical & \(\times 10\) quantum-time & \(\times 10^{-3}\) classical-time \\
        \hline
        $N^4$ HF      & 2044   & 2044   & $>2050$   & $>2050$ \\
        $N^5$ MP2     & 2038   & 2036   & 2039      & 2041    \\
        $N^6$ CCSD    & 2036   & 2034   & 2036      & 2037    \\
        $N^7$ CCSD(T) & 2034   & 2032   & 2035      & 2036    \\
        $4^N$ FCI     & 2031   & 2029   & 2032      & 2032    \\
        \hline
        \vspace{1pt}
    \end{tabular}
    }
    \footnotesize{Variation of our predicted year of disruption based on changes in logical qubit requirements and quantum or classical constants.}
\end{table}

\subsection{Limitations and Further Work}\label{Limitations and Further Work}
In this investigation, we have tried to do a much broader comparison of quantum and classical algorithms. This means we have had to make broad generalizations about chemistry methods and quantum computing. We have tried to outline these limitations in our work. We are particularly aware of the limited information we have on the algorithmic constants. Our current estimates are meant to capture reasonable scenarios not meant to capture the diversity of implementations. We also have made broad generalizations about quantum computing focusing in on superconducting qubits under IBM's roadmap as a tractable approach. Further, there is still little data on the cost and overhead of quantum computers so we have had to make rough estimates based on the best information we have, see Appendix ~\ref{How Fast Are Quantum Computers?}. 

Despite the breadth of our study, there is still much more work to be done comparing quantum and classical chemistry methods. We have primarily focused on ground state estimation in quantum chemistry. This is one of the oldest and most important areas of computational chemistry. However, quantum computers have utility in other domains of chemistry and physical simulations. An interesting case is that of simulating electron dynamics where quantum computers have a particularly strong advantage in the Warm Dense Matter Regime \cite{Babbush_2023}. We have also neglected the analysis of other quantum methods, such as variational methods in quantum chemistry (see \ref{sec:Quantum Chemistry Algorithms on Quantum Computers}).

Another important topic we have neglected in our discussion is the role of new classical technologies like neural networks in the future of quantum chemistry. These systems are already replacing detailed physics based simulation. Tools such as eSEN \cite{fu2025learning} are able to predict interatomic potentials and perform well on the Matbench Discovery benchmark. We can see a world where these systems replace many classical and quantum algorithms. Yet, another strong approach developed in current systems is complementary where neural network and first-principled approaches (quantum and classical) trade off each other \cite{techreview2024ai}. Quantum computers can also benefit from AI, and hope lies in quantum machine learning for quantum systems. In these cases, quantum computers have a theoretical asymptotic exponential advantage in learning from quantum experiments over classical neural networks \cite{huang2022quantum}.

\section{Conclusion}\label{Conclusion}
We have gained insights into particularly important fault-tolerant methods to disrupt computational chemistry. We've investigated the complexities and limitations of quantum computers and quantum chemistry algorithms. Overall, we see that quantum methods have the potential to disrupt a wide variety of high-accuracy classical computational methods. In particular, improved asymptotic times are an incredibly important driver of progress. In addition, the 2030s look to be an especially eventful time for quantum computers. In the early 2030s we will see some of the first industry-relevant applications of quantum technology. Nevertheless, with current algorithms and future hardware overheads it looks unlikely that quantum algorithms will replace the majority of calculations done in quantum chemistry. Still, there is significant room for optimism. We see significant progress in both algorithmic constants and algorithmic time complexity. Quantum hardware may also have a much larger advantage with new error correction techniques, quantum hardware beyond superconducting qubits, and the possible stagnation of classical computing hardware. 

We hope that quantum computers can realize their potential and help us gain insight into the quantum systems that underpin our world. Our final forecast is that the era of quantum computing is just getting started. We do not view our predictions as immutable facts but instead as a call to action for progress in a new computing paradigm.

\appendix


\subsection{How Fast Are Quantum Computers?}\label{How Fast Are Quantum Computers?}

\begin{table*}[!h]
\centering
\caption{Summary of Flop-Adjusted Constants for CCSD/CCSD(T) Benchmarks with Citations}
\label{tab:classical_benchmark}
\begin{tabular}{lccccccc}
\hline
\textbf{Hardware} & \textbf{Method} & \textbf{\(N\)} & \textbf{\(T\) (s)} & \textbf{\(P\) (flops/s)} & time-complexity & \(\boldsymbol{c_{\rm alg}=T\cdot P/N^p}\) & \textbf{Citation} \\
\hline
8 V100 GPUs & CCSD & 966 &  \(960\)  &  \(9.8\times10^{13}\) & $O(N^6)$ & 0.12 &\cite{Yue2020ccsd_gpu} \\
6 Tesla K40 & CCSD(T) & 315 &  \(61200\) & \(8.4\times10^{12}\) & $O(N^7)$ & 1.67 & \cite{nvidia2016quantum} \\
\hline
\end{tabular}
\\[0.5ex]
\footnotesize{We estimate the algorithmic constant overhead for CCSD, and CCSD(T) separate from the hardware speed of floating point operations. This involves looking at the ratio between the estimate of the FLOPs required to do that operation divided by the estimate based on the asymptotic formula. N refers to  basis functions.}
\end{table*}

\begin{table*}[!h]
\centering
\caption{Comparison of Logical Qubit and T-gate Estimates for QPE}
\label{tab:Logical qubit and T-gate estimats for QPE}
\begin{tabular}{lccccccc}
\hline
\textbf{Chemical System} & \textbf{Naive-T-gate Estimate} & \textbf{\(N\)} & \textbf{Actual T-gates} & \textbf{logical qubits} & \textbf{QPE time complexity}  & \textbf{Citation} \\
\hline
Beta Amyloid (5+Cu) fragment         & $2.6 \times 10^{14}$     & 192  & \(1.17 \times 10^{14}\)             & \(4728\) & $O(N^{5}/\epsilon)$ &  \cite{otten2024quantum} \\
FeMoCo &  $2.0 \times 10^{7}$  & 54 & $3.41 \times 10^{8}$  & 2142 & $ ~ O(N^3/\epsilon)$ & \cite{low2025fastquantumsimulationelectronic} \\

\hline
\end{tabular}
\\[0.5ex]
\footnotesize{We estimate T-gates in our model naively. We make T-gate estimates by taking the number of basis functions, exponentiation to the relevant asymptotic exponent ($N^5$ or $N^3$ and multiplying by $1/\epsilon$). We assume an accuracy of $\epsilon = 10^{-3}$. N refers to basis functions for that asymptotic expression. For many recent QPE implementations, our naive estimates are accurate enough. \cite{low2025fastquantumsimulationelectronic} used a method with scaling $O(N^2) - O(N^4)$ we chose or default $O(N^3)$ estimate here.}
\end{table*}

Our study aims to compare the efficacy of quantum and classical algorithms under equivalent resource constraints, i.e., same time-frame and budget. Due to the significant cost disparity between quantum and classical computing, our estimates allow for a considerably larger number of classical operations. We base this on cloud prices for a Rigetti quantum computer which are approximately $1.3\$$ per second \cite{azure_quantum_pricing}, and IBM's 27-qubit Falcon processor, which has a price of $1.60\$$ per second \cite{helsel2022ibm}. This is in comparison to the price per second of an NVIDIA H100 GPU, which has a cloud price of approximately $0.001\$$ dollars per second \cite{aws_ec2_capacityblocks_pricing}. 

We adopt the initial classical and quantum speed difference as established in \cite{choi2023quantumtortoiseclassicalhare}. By default, we set the classical clock speed at 5 GHz and the quantum clock speed at 2 MHz. The error correction overhead is assumed to introduce a slowdown on the order of $10^{2}$ for each logical operation \cite{choi2023quantumtortoiseclassicalhare}. Combining these two slowdowns leads to a naive quantum slowdown of $10^{2} \cdot 10^{3} = 10^{5}$ relative to a classical processor. In other words, a quantum computer can perform about $10^{6}$ physical operations and $10^{4}$ logical operations per second. 

\subsubsection{Parallelization Overhead}
Given the cost and speed information above, a superconducting quantum computer can implement approximately $10^6$ serial physical gate operations per second with $1 \$$. However, a quantum computer can implement gate operations on each qubit line at a given instant, and quantum algorithms can utilize some of this parallelization, so we estimate that a quantum computer can do almost $10^{7}$ physical gate operations or $10^5$ logical operations. Given the lower per second cost of the Nvidia H100 GPU, we can run approximately $10^{3}$ H100 GPUs in one second for one dollar (this would be $3600 \$$ per hour). A single H100 GPU is capable of $\approx 10^{15}$ floating point operations per second \cite{nvidia_h100_tensor_core_gpu_datasheet} classically with full utilization. This means that with $1 \$$, it is possible to do $10^{18}$ floating point operations per second. Compared to parallel quantum logical operations, this indicates a $10^{13}$ overhead. This overhead is due in part to the $10^{5}$ naive speed overhead. We attribute the remaining $10^{8}$ overhead to increased parallelism caused by the quantum-classical price difference, i.e., you can run $10^{8}$ more processes on a quantum computer than a classical computer at the same cost.

\subsection{Orbital to Atom Ratio}
\label{sec:Orbital to Atom Ratio}
Basis function (orbital basis function) counts are not an intuitive way to view the results for non-chemists. We offer a heuristic rough estimate of the ratio between basis functions/orbitals here to convert our results to atomic system size.
The number of orbitals necessary for a chemical system is highly contingent on the type of molecule and the accuracy needed. For instance, if plane wave basis functions are used, many more basis functions may be necessary to obtain accurate results for localized molecules \cite{Berry_2019}. On the other hand, asymptotic analyses for FCI often assume a minimal basis set, where there is exactly one basis function corresponding to each occupied atomic orbital \cite{Field2017}.

To get a rough estimate for a standard basis set in quantum chemistry, we look at the orbital-to-atom ratio for FeMoCo (an important chemical in quantum simulation).

We can estimate the number of orbitals of FeMoco (Fe$_7$MoS$_9$C) using the mixed 6-31G + LANL2DZ basis set, where LANL2DZ will be employed for the transition metals. Explicitly, 6-31G is used for carbon and sulfur atoms, and LANL2DZ for iron and molybdenum.

In such a scheme, we obtain:
\begin{itemize}
    \item Fe: 7 atoms $\times$ 22 orbitals/atom = 154 orbitals
    \item Mo: 1 atom $\times$ 22 orbitals/atom = 22 orbitals
    \item S: 9 atoms $\times$ 13 orbitals/atom = 117 orbitals
    \item C: 1 atom $\times$ 9 orbitals/atom = 9 orbitals
\end{itemize}
   
which gives a total number of orbitals of 302, and an orbital-to-atom ratio of 16.8.

A slightly lower estimate might look at modeling
hydrocarbon chains with 6-31G basis set (e.g., CH$_3$–(CH$_2$)$_n$–CH$_3$): \\
For each CH$_2$ unit:
\begin{itemize}
    \item C: 9 orbitals
    \item 2 H: $2 \times 2 = 4$ orbitals
\end{itemize}
Which gives an orbital-to-atom ratio of :
$\frac{13}{3} \approx 4.3$

\bibliographystyle{IEEEtran}
\bibliography{IEEErefs}

\end{document}